\documentclass[pdflatex,sn-nature]{sn-jnl}

\usepackage{graphicx}%
\usepackage{multirow}%
\usepackage{amsmath,amssymb,amsfonts}%
\usepackage{amsthm}%
\usepackage{mathrsfs}%
\usepackage[title]{appendix}%
\usepackage{xcolor}%
\usepackage{textcomp}%
\usepackage{manyfoot}%
\usepackage{booktabs}%
\usepackage{algorithm}%
\usepackage{algorithmicx}%
\usepackage{algpseudocode}%
\usepackage{listings}%
\usepackage[ngerman]{}
\begin{document}

\title{Optical and spin properties of nitrogen vacancy centers formed along the tracks of high energy heavy ions}

\author*[1]{Wei Liu}\email{weiliu01@lbl.gov}
\author[2]{Aleksi A. M. Leino}
\author[1]{Arun Persaud}
\author[1]{Qing Ji}
\author[1]{Kaushalya Jhuria}
\author[3]{Edward S. Barnard}
\author[3]{Shaul Aloni}
\author[4,5]{Christina Trautmann}
\author[4,6]{Marilena Tomut }
\author[1,7]{Ralf Wunderlich}
\author[2]{Chloé Nozais}
\author[8]{Saahit Mogan} \author[8]{Hunter Ocker} \author[8]{ Nishanth Anand} \author[8]{Zhao Hao}

\affil*[1]{Accelerator Technology and Applied Physics Division, Lawrence Berkeley National Laboratory, Berkeley, USA}

\affil[2]{Helsinki Institute of Physics, Department of Physics, University of Helsinki, Helsinki, Finland}

\affil[3]{Molecular Foundry, Lawrence Berkeley National Laboratory, Berkeley, USA}

\affil[4]{GSI Helmholtzzentrum f{\"u}r Schwerionenforschung, Darmstadt, Germany}
\affil[5]{Technische Universit{\"a}t Darmstadt, Institute of Materials Science, Darmstadt, Germany}
\affil[6]{Institute of Materials Physics, WWU Münster, Münste, Germany}
\affil[7]{Faculty of Physics and Earth System Sciences, Felix Bloch Institute for Solid State Physics, Applied Quantum Systems, Leipzig University, Leipzig, Germany}
\affil[8]{Earth and Environmental Sciences, Lawrence Berkeley National Laboratory, Berkeley, USA}
\author[2]{Flyura Djurabekova}
\author*[1]{Thomas Schenkel}\email{T\_Schenkel@lbl.gov}

\abstract
{Exposure of matter to high energy, heavy ions induces defects along the trajectories of the ions through electronic and nuclear energy loss processes. Defects, including color centers, can recombine or form along latent damage tracks in many materials, such as insulators and semiconductors. Latent tracks in diamond were only recently observed. Here, we report on color center formation in diamond along the latent tracks of 1 GeV gold and uranium ions. Using depth-resolved photoluminescence, we observe direct formation of single vacancy related color centers (GR1-centers) along the ion tracks. Mobile vacancies can form NV-centers with native nitrogen atoms during thermal annealing. Molecular dynamics simulations show that isolated vacancies and vacancy clusters form through electronic stopping processes, leading to color center formation along ion trajectories from the sample surface to a depth of about 25 microns. We further report on the creation of individually isolated quasi-1D chains of NV-centers by using 1 GeV Au ions with a dilute fluence. The individual 1D NV-chains appear as isolated bright luminescence strings and present competitive electron spin properties compared to a background of NV-centers. Such spin textures can be explored as building blocks for applications in quantum information processing and quantum sensing.}

\keywords{Swift heavy ions, Diamond NV-centers, GR1-centers, ODMR, PL}

\maketitle

\section*{Introduction}
High energy, heavy ions (or swift heavy ions, SHI), deposit tens of keV of energy per nanometer when they impinge on materials \cite{lang2020fundamental}. The intense local excitation leads to the formation of vacancies and interstitials along ion trajectories. Many of these defects quickly recombine, while others form stable defect centers, including optically active defects, or color centers. The track dimensions of SHI e. g. in polymers or semiconductors have unique nm-scale alignment over tens of microns. In diamond, latent tracks of damage centers were only very recently observed \cite{amekura2024latent}. Recently, evidence for the formation of ensemble quasi-1D chains of NV-centers in diamond formed by SHI has been reported in diamonds that contained 100 ppm of nitrogen \cite{lake2021direct}. Here, we report on the formation process of NV-centers in quasi-1D chains and 2D sheets using 1 GeV gold and uranium ions and single crystal diamonds that contain 1 ppm nitrogen. At this lower nitrogen concentration, the optical signal from background NV-centers is reduced, thereby enabling us to unravel the role of single vacancy centers (GR1) along ion trajectories in the formation of spin textures with NV-centers. We further create individually isolated quasi-1D chains of NV-centers in the diamond ($<$200 ppm of nitrogen) by using 1 GeV Au ions with a dilute fluence. The individual 1D NV-chains appearing as isolated bright luminescence strings, which indicates the presence of densely coupled NV-centers created along a single ion trajectories.

Negatively charged nitrogen-vacancy (NV$^{-}$) color centers in diamond possess an optical accessible spin-1 triplet ground state with up to millisecond longitudinal relaxation time $T_{1}$ and coherence time $T_{2}$ at room temperature \cite{rondin2014magnetometry}. Optical selective transitions enable utilizing NV$^{-}$ centers for quantum sensing via optical detected magnetic resonance (ODMR), often with superior sensitivity compared to conventional magnetometers \cite{barry2020sensitivity}. Thanks to the technical simplicity, radiation robustness, chemical inertness, and nanoscale geometry, diamond-NV sensing can be deployed in harsh radiation, biochemical, and geoscience-related environments \cite{zhang2021toward,fu2020sensitive,bakhshandeh2022quantum,liu2023nanothermometry}. The control of multiple coupled NV$^{-}$ centers \cite{dolde2013room,jarmola2012temperature,bauch2018ultralong} is an essential prerequisite for realizing a variety of functionalities of diamond-NV and unitary fidelity of quantum protocol operations \cite{taminiau2014universal,waldherr2014quantum,rong2015experimental,wang2015quantum,pezzagna2021quantum, hensen2015loophole}. 

Recent studies indicate self-aligned quasi-1D chains of coupled NV$^{-}$ centers along a length of several tens of microns \cite{lake2021direct}. This effect is promising for the development of a novel type of quantum register and a building block for NV-based quantum information processing \cite{tsuji2022high, greengard2021qubit,hamiltonheterogeneous}. Such quasi-1D chain of NV$^{-}$ centers can be realized by swift heavy ions (SHI) of sufficiently high electronic stopping power such as e.g. 1-2 GeV gold or uranium ions irradiation of single crystal diamonds. The energy deposition along the ion trajectory of several tens of keV per nm leads to the formation of vacancy centers and the conversion of native nitrogen atoms to NV-centers. Adjusting the nitrogen concentration, and the energy and species of the ions allow engineering the longitudinal NV$^{-}$ spacing on the few nanometer scale and resulting spin chains on the length scale of a tens of microns \cite{lake2021direct,onoda2017diffusion,onoda2015new}. Quasi-1D spins chains and 2D sheets of aligned NV$^{-}$ centers enables studies of Ising model spin dynamics and explorations of spin transport and spin registers \cite{bose2003quantum}. Formation of GR1-centers and NV-centers in diamond by SHI also supports the development of diamond-based single ion track detectors \cite{onoda2015new,akselrod2018fluorescent} and methods for directional detection of hypothetical highly energetic dark matter candidates \cite{marshall2021directional}. Further, the optical readout of NV-centers along ion tracks supports experimental benchmarking of molecular dynamics (MD) simulations of the interaction between SHI and the diamond lattice \cite{liu2020molecular}.

In this study, we present insight into GR1 and NV-center generation along the tracks of SHI in diamond and quantify the resulting spin properties of NV-centers. Diamond with $<$1 ppm N density was chosen in order to suppress decoherence induced by the nitrogen spin bath \cite{bauch2018ultralong} and to enable the observation of GR1-centers in the spectral range that overlaps with that from NV$^{-}$ centers. By using confocal laser scanning fluorescence microscopy \cite{lake2021direct}, we characterized GR1, NV$^{0}$ and NV$^{-}$ centers and the conversion from GR1's to NV-centers. We show that the NV-center formation dynamics by SHI in diamonds with relatively low nitrogen content follows a two-step process of vacancy formation followed by capture of mobile vacancies by substitutional nitrogen atoms. The SHI induced vacancies can act as optically active GR1-centers, when presented as individual isolated neutral vacancies. Vacancies form predominantly along the ion path where the electronic stopping power is highest. They can further combine with the nearby nitrogen atoms to form NV-centers during the cools down of the excited track zone. NV$^{-}$ center formation is further enhanced during thermal annealing after SHI irradiations. The micrometer-resolved optical analysis on the NV-Center formation provides experimental benchmarking data for dynamic Monte Carlo simulations of ion track structures and micro-dosimetric models. To further create coupled NV-centers spin chain, we chose diamond with $<$200 ppm N density, which allows created dense coupled NV-centers with spacing in the range of a few nm along single ion trajectories. Individual quasi-1D chains of NV-centers were created by using 1 GeV Au ions with a dilute fluence and post annealing. We probe the spin properties of SHI-induced NV$^{-}$ via ODMR, which shows that SHI-induced NV$^{-}$ centers along quasi-1D chains or in thin sheets with length of about 30 $\mu$m can be used for applications in high-sensitivity magnetometry and for studies of spin textures in diamonds. Irradiation with SHI is a method for NV-center formation, complementary to more common irradiation with MeV electrons and protons \cite{acosta2009diamonds}, with the unique feature of alignment of NV-centers along the (mostly) straight SHI trajectories in quasi-1D chains with a widths of a few nm and a length of tens of microns.

\section*{Results and Discussion}
\subsection*{Optical properties of color centers formed along the SHI trajectories}
Type IIa diamonds (supplier: Element 6) were used in this study with about 1 ppm nitrogen introduced during the chemical vapor deposition growth. In a first step, we implanted erbium ions into diamond samples (1$\times$10$^{13}$ cm$^{-2}$ 180 keV). Er ion implantation \cite{cajzl2018erbium} led to formation of NV-centers near the sample surface, within the top 100 nm as estimated by the stopping and range of ions in matter modeling (SRIM-2008) \cite{SRIM}. Emission from NV-centers near the surface serves as a reference of the sample surface for depth dependent PL measurements. We did not observe optical emission from Er atoms or signatures of NV-Er coupling in optical spectra or in ODMR. Next, pre-implanted samples were irradiated at the linear accelerator UNILAC at GSI Helmholtzzentrum (Darmstadt, Germany) using 1.1 GeV U ions (fluence of 1$\times$ 10$^{12}$ U ions/cm$^{2}$) and 0.95 GeV Au ions (2.2$\times$10$^{12}$ Au ions/cm$^{2}$), labeled as sample A and B, respectively. The electronic stopping power at the sample surface and the range in diamond are 49 keV/nm and 30 µm for U ions and 40 keV/nm and 30 µm for Au ions (estimated using SRIM). These rather similar values of both irradiation conditions allow us to observe common SHI related effects and trends. The samples were covered by a thick honey-comb shaped mask with millimeter sized openings (90\% transparency), which allows direct comparison between irradiated and nonirradiated regions during the PL measurements. To enhance the formation of NV$^{-}$ centers, the Au-ion irradiated sample B was thermally annealed in a two-step process and the optical properties were compared after the 1st and the 2nd annealing step. The 1st annealing step was for 1 hour at 800 \textdegree C in vacuum (10$^{-6}$ mbar), the 2nd annealing step was for 1 hour at 1000 \textdegree C in argon atmosphere. After the two annealing steps, we performed secondary ion mass spectrometry (SIMS) measurements on sample B after the thermal annealing steps (See figure S01 in the Supplementary Material). SIMS depth profiles show accumulation of nitrogen to concentrations of up to 1$\times$10$^{19}$ cm$^{-3}$ ($\sim $ 60 ppm) within a surface layer of 100 nm depth and a concentration of ~3$\times$10$^{16}$ cm$^{-3}$ (0.2 ppm) for larger depths. This value is consistent with the $<$ 1 ppm indicated by the supplier, but may also be affected by the absolute calibration in SIMS measurements. The Er profile is distributed within the top 300 nm with an apparent areal density of 4.1$\times$10$^{12}$ cm$^{-2}$. We note that the absolute concentration numbers are possibly affected by the SIMS calibration.

We performed depth-dependent PL on U-ion irradiated samples (sample A, without thermal annealing) to characterize the interaction between nitrogen and SHI-formed vacancies. 
\begin{figure}[!]
\centering
\includegraphics[width=1\columnwidth]{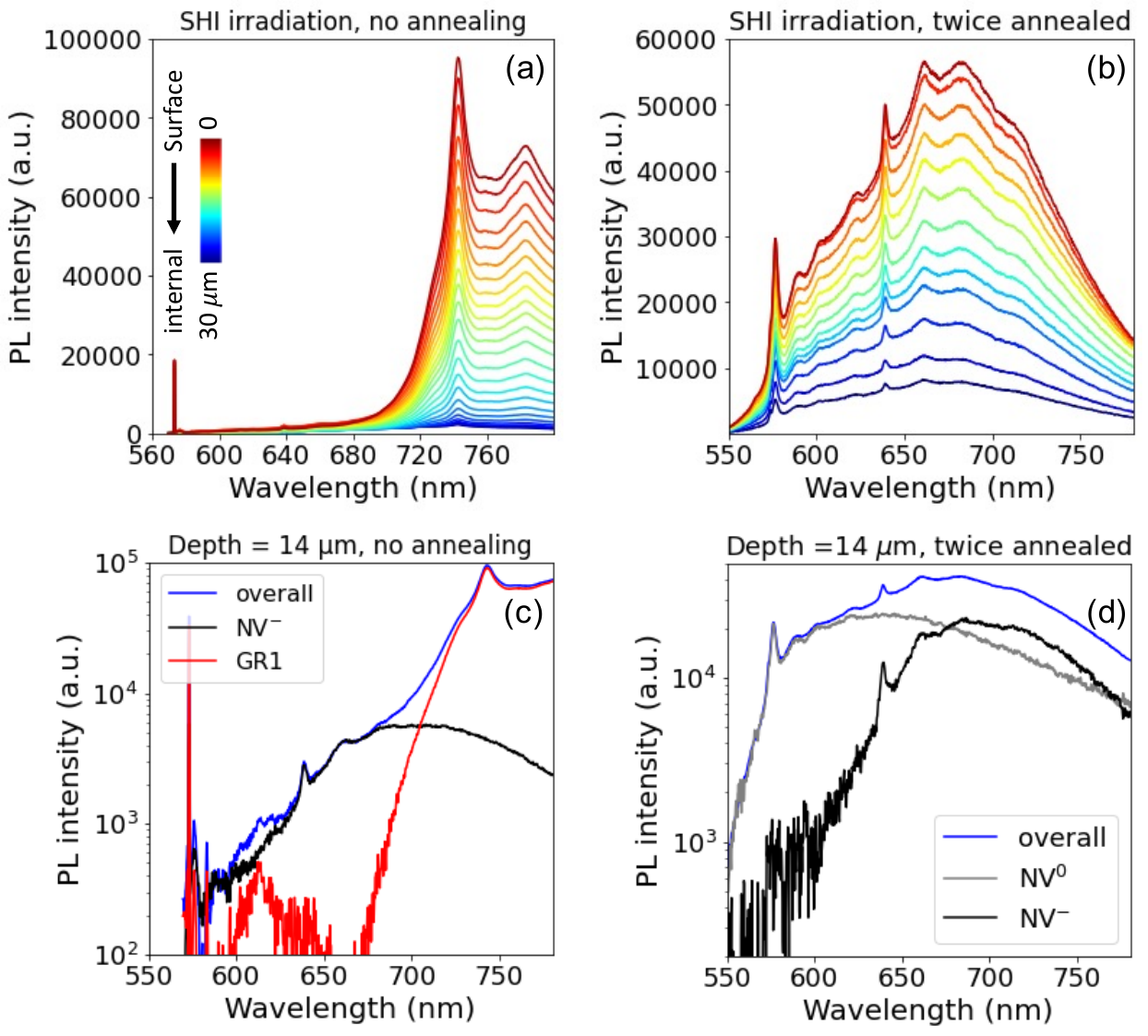}
\caption{\textbf{Depth-resolved PL spectra of color centers formed along the SHI trajectories.} (a) Color center PL spectra of sample A (no annealing) as a function of depth from the sample surface to a depth of 30 $\mu$m in the area irradiated with 1$\times$ 10$^{12}$ U ions/cm$^{2}$ fluence of 1.1 GeV energy. (b) Color center PL spectra of sample B (after two annealing steps) as a function of depth from the sample surface to a depth of 30 $\mu$m in the area irradiated with 2.2$\times$ 10$^{12}$ Au ions/cm$^{2}$ fluence of 0.95 GeV energy. (c) Deconvolution of the PL spectrum (blue) into NV$^{-}$ and GR1 luminescence component probed in the SHI irradiated area of sample A at depth = 14 $\mu$m (roughly half the SHI range). The red and black spectra show the deconvoluted spectrum GR1 and NV$^{-}$, respectively. (d) Deconvolution of sample B probed in the SHI irradiated area at a depth of 14 $\mu$m after annealing. The black and grey spectrum shows the decomposed respective data for NV$^{-}$ and NV$^{0}$ centers.}
\label{fig0}
\end{figure} 
Figure \ref{fig0} (a) shows the PL spectra as a function of the probe depth (refractive index corrected \cite{lake2021direct}) from the irradiated surface (red) to a depth of 30 $\mu$m (marked as blue). In the area absence of U-ion irradiation, we observe a weak GR1 zero phonon line (ZPL) signal (figure S02 in the Supplementary Material) that is visible near the sample surface, within the implantation range of the Er ions. When probing deeper regions, the spectra are dominated by a weak emission from native NV-centers that are present in the as-received sample. The luminescence of GR1-centers is known to originated from individual neutral vacancies \cite{subedi2021spectroscopy}, which evidences that the SHIs generate individual isolated vacancies along their trajectories by displacing carbon (or nitrogen) atoms from their lattice position. We also observe weak luminescence from NV$^{-}$ and NV$^{0}$ center at the ZPLs of 637 nm and 573 nm, respectively. This phenomenon is different from our previous experiment where preferentially NV$^{-}$ centers were formed in diamond that contained 100 ppm nitrogen and that had been irradiated with 1 GeV ions (10$^{12}$ cm$^{-2}$) \cite{lake2021direct}. The major difference in the present experiment is the more than 2 orders of magnitude reduced intrinsic nitrogen concentration of ~1 ppm, while the ion irradiation conditions are similar. Figure \ref{fig0}(b) shows the depth-dependent PL spectrum of Au-ion irradiated areas of sample B after thermal annealing. Here, the spectra are dominated by the typical ZPL of NV$^{0}$ at 573 nm and NV$^{-}$ at 637 nm, while spectral signatures of GR1-centers now absent. The annihilation of GR1 luminescence in sample B after the annealing shows that the SHI-irradiation induced vacancies either combine with nitrogen atoms to form NV-centers or annihilated e.g. via recombination with carbon interstitials. 

To quantify the PL intensity and analyze the conversion between GR1 and NV$^{-}$ along the SHI trajectories, Fig. \ref{fig0} (c) shows an example of the deconvolution of the spectra into NV$^{-}$ and GR1 luminescence components of sample A (at depth = 14 $\mu$m in the SHI irradiated area) \cite{jeske2017stimulated,sola2019electron}. The deconvolution approach is described in the Supplementary Material. The GR1 PL intensity is more than one order of magnitude higher than the NV$^{-}$ intensity after exposure to SHI and before annealing. This shows that for diamond with a low nitrogen concentration (nominal 1 ppm) the share of SHI-induced vacancies that combing with nitrogen and form NV$^{-}$ centers is small. This is in contrast to earlier experiments, where strong NV$^{-}$ center emission was observed directly after SHI irradiations and before thermal annealing \cite{lake2021direct}. At the lower nitrogen concentration, SHI induced vacancies have to move farther to find a nitrogen atom for NV-center formation. At a nitrogen concentration of 100 ppm, the average distance between N atoms is about 5 nm, while it is ~20 nm at ~1 ppm. Vacancies can diffuse during track cool down or thermal annealing and form NV-centers if enough N atoms are present near the ion trajectory. 

As shown in figure \ref{fig0}(d), in the SHI irradiated area, the ratio of the NV$^{-}$ spectral area to that of NV$^{0}$ is enhanced after the 2nd thermal annealing step compared to the 1st annealing step (see figure S03 in the Supplementary Material). NV$^{0}$ centers form first during thermal annealing and the charge state balance shifts to NV$^{-}$ via NV$^{0}$ + $e^{-}$ $\rightleftharpoons$ NV$^{-}$ during longer annealing times, when vacancy centers dissolve and electrons from the relatively sparse N density can be picked up by thermally activated charge transfer \cite{luhmann2021charge}. Performing a 2nd annealing step allows us to thermally drive this process by further promoting the ionization of nitrogen donors and to transfer the electron charge to the site of NV$^{0}$ centers, consequently converting NV$^{0}$ to NV$^{-}$.

\subsection*{MD simulation of the interaction between SHI and the diamond lattice: an insight into PL measurement}
\begin{figure}[h]
\centering
\includegraphics[width=\columnwidth]{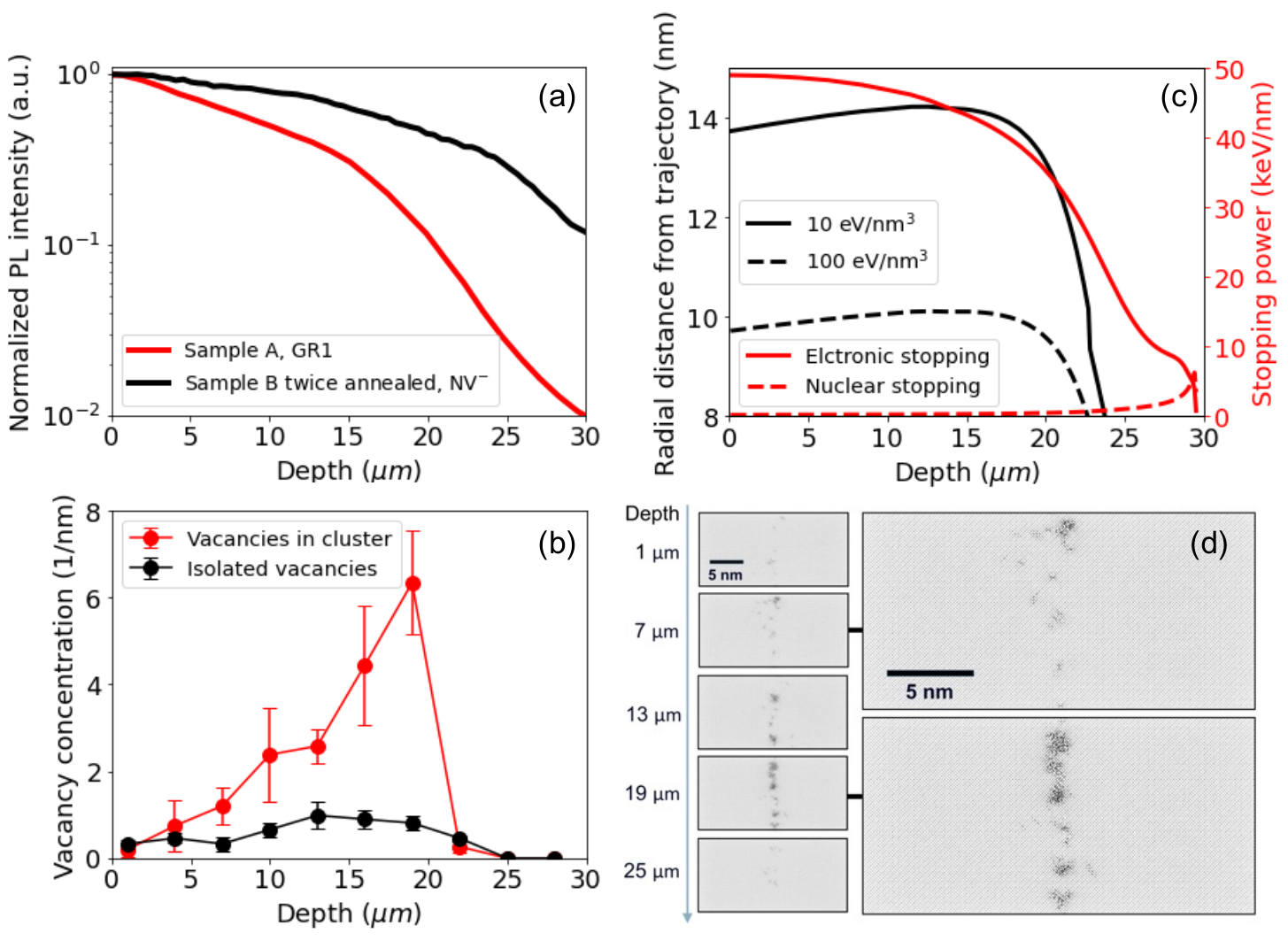}
\caption{\textbf{MD simulation insight into PL measurement on the interaction between SHI and the diamond lattice.} (a) Comparison of depth-resolved normalized PL intensity (logarithmic scale) of GR1 defect in sample A (no annealing) and NV$^{-}$ defect in the annealed sample B. (b) The simulated concentrations of vacancies in the form of isolated vacancies and vacancy clusters along the ion trajectory from electronic energy loss processes according to MD simulations. The contribution from elastic collisions as estimated by SRIM is about 0.5 vacancies/nm in the top 15 microns. (c) Electronic and nuclear stopping powers along the ion trajectory of 1.1 GeV U ions in diamond (right axis). The left axis shows the contour lines of the initial energy density of 10 eV/nm$^{3}$ and 100 eV/nm$^{3}$ after ion impact in the electronic subsystem as estimated from the delta-ray dose formulas (See supplementary material). (d) Visualization of the MD simulation cells at different depth so that atoms are drawn as gray dots. The arrow on the left shows the ion propagation direction. The left two subfigures show the zoom-in area with defect clusters at 7 $\mu$m and 19 $\mu$m depth.}
\label{fig1}
\end{figure}

Figure \ref{fig1} (a) compares depth profiles of the normalized PL intensity of the GR1 luminescence of sample A with the emission of NV$^{-}$ centers that is dominant in the annealed sample B. The two PL intensity profiles show a similar plateau up to a depth of 15 $\mu$m. With the electronic stopping range from 20 to 30 $\mu$m depth, the GR1 PL intensity in sample A dropping more drastically than the NV$^{-}$ PL in sample B, the GR1 PL intensity between sample surface and the regime of end-of-range ion track differs by two orders of magnitude. The drastically decreased of GR1 PL intensity at the end of electronic stopping range can be related to the clustering of vacancies, which can render the centers optically inactive. The formation of vacancy complexes can be further enhanced by nuclear stopping process at the end of the ion path \cite{lake2021direct}. In contrast, in sample B, the reduction of PL intensity of NV$^{-}$ center emission from the sample surface to the end of the ion range is less than 1 order of magnitude. We infer that the termal annealing converts the additional vacancies and substitutional nitrogen atoms into NV$^{-}$ centers near the end of the ion range. These results show that SHI efficiently introduce vacancies along their trajectories in areas of high electronic stopping, as well as at the end of their range where elastic collisions dominate. It also indicates that the NV-center formation dynamics by SHI is governed by a two-step process that depends on the available nitrogen atoms around the ion trajectory. The SHI-induced vacancies can be in the form of individual neutral vacancies or further combine with the neighboring N to form NV-centers. The strong GR1 luminescence introduced directly by SHI irradiation (without thermal annealing) as well as damage repair and re-crystallization via thermal annealing are important features for applications of diamond-based radiation detectors, such as fluorescent nuclear track detectors for extreme radiation environments \cite{onoda2017diffusion,onoda2015new}.

To gain insight into vacancy formation resulting from electronic stopping processes for 1.1 GeV U ion, we performed two-temperature MD simulations in pure diamond using the Tersoff potential \cite{tersoff1989modeling}. The detailed model is described in Ref.\cite{amekura2024latent} and also in the Supplementary Material. The simulations here exclusively consider the electronic component, omitting nuclear stopping power effects. To circumvent the computation limits, the 30 $\mu$m long ion trajectory was segmented into ten small simulations depicting small slabs along the path, each with dimensions of 23$\times$23$\times$11 nm$^{3}$ (XYZ). The trajectory is centered in X and Y directions and penetrates through the cell along the Z axis. Figure \ref{fig1}(b) presents vacancies concentrations along the trajectory determined using the Wigner-Seitz analysis. The plot shows that the concentration of isolated vacancies (Voronoi cells) slightly increase from 0.5 nm$^{-1}$ to 1 nm$^{-1}$, as the ion track extends from surface to the internal depth of 20 $\mu$m. In comparison, the vacancies density in the form of clusters (empty Voronoi cells connected to other cells) increases drastically from 0 to 7 nm$^{-1}$ from the surface to a depth of 20 $\mu$m). It is surprising to witness this dramatic increase of vacancy clustering in the range of electronic stropping, while the electronic stopping power decreases monotonically from the surface as shown in \ref{fig1}(c). We tentatively ascribe the enhanced vacancy clustering near 20 $\mu$m to the ion velocity effect. Our simulation of the delta-ray radial dose distribution in figure \ref{fig1}(c) reveals that the initial energy density near the trajectory increases due to the decreased velocity of the ion \cite{wang1994se} until about 20 $\mu$m. Higher energy densities result in stronger defect production. Focusing on the role of electronic stopping processes, we neglect the contributions to vacancy production from elastic collisions. From SRIM we estimate the rate of vacancy formation to be $\sim$0.5 vacancies/nm from the surface to a depth of ~20 microns. 

Figure \ref{fig1}(d) shows atomic disorders along an ion trajectory from MD simulations at different depth, where atoms are drawn as gray dots. The details of isolated defects and clusters can be seen in the enlarged visualization of the cell at 7 $\mu$m and 19 $\mu$m depth. The MD simulation results clearly reveal that close to the surface of $<$ 10 $\mu$m, the damage consists of both isolated interstitial-vacancy (Frenkel) pairs and small defect clusters. The size of the defect clusters tends to gradually increase within a depth between 13 $\mu$m and 20 $\mu$m. In particular, we observe graphitization of the diamond structure showing amorphous defect clusters near 20 $\mu$m depth. In the 20-25 $\mu$m range, there is a sharp decrease in the defect concentration, until no defects form at depths greater than 25 $\mu$m. Such a velocity effect with graphitization and a sharp decrease of vacancy concentration in the 20-25 $\mu$m range can well explain the experimentally observed drastic drop of GR1 emission in this depth range. However, as mentioned above, the simulation does not account for the nuclear stopping power, due to the calculation complexity. As ions slow down, the nuclear stopping power component becomes increasingly significant compared to electronic stopping leading to the formation of defects at the end of the ion ranges, potentially leading to further vacancy defects. Potential synergies of elastic and inelastic energy loss processes on defect kinetics will be explored in future studies \cite{schenkel1998synergy,nuckols2021effects}. 

\subsection*{Electron spin properties of ensemble NV$^{-}$ centers with deep subsurface placement formed by high fluence SHI irradiation}
We performed ODMR measurements to characterize the electron spin (m$_{s}$ = +1) properties of ensemble NV$^{-}$ centers with deep subsurface placement formed by high fluence SHI irradiation after thermal annealing (sample B).
\begin{figure}[h]
\centering
\includegraphics[width=\columnwidth]{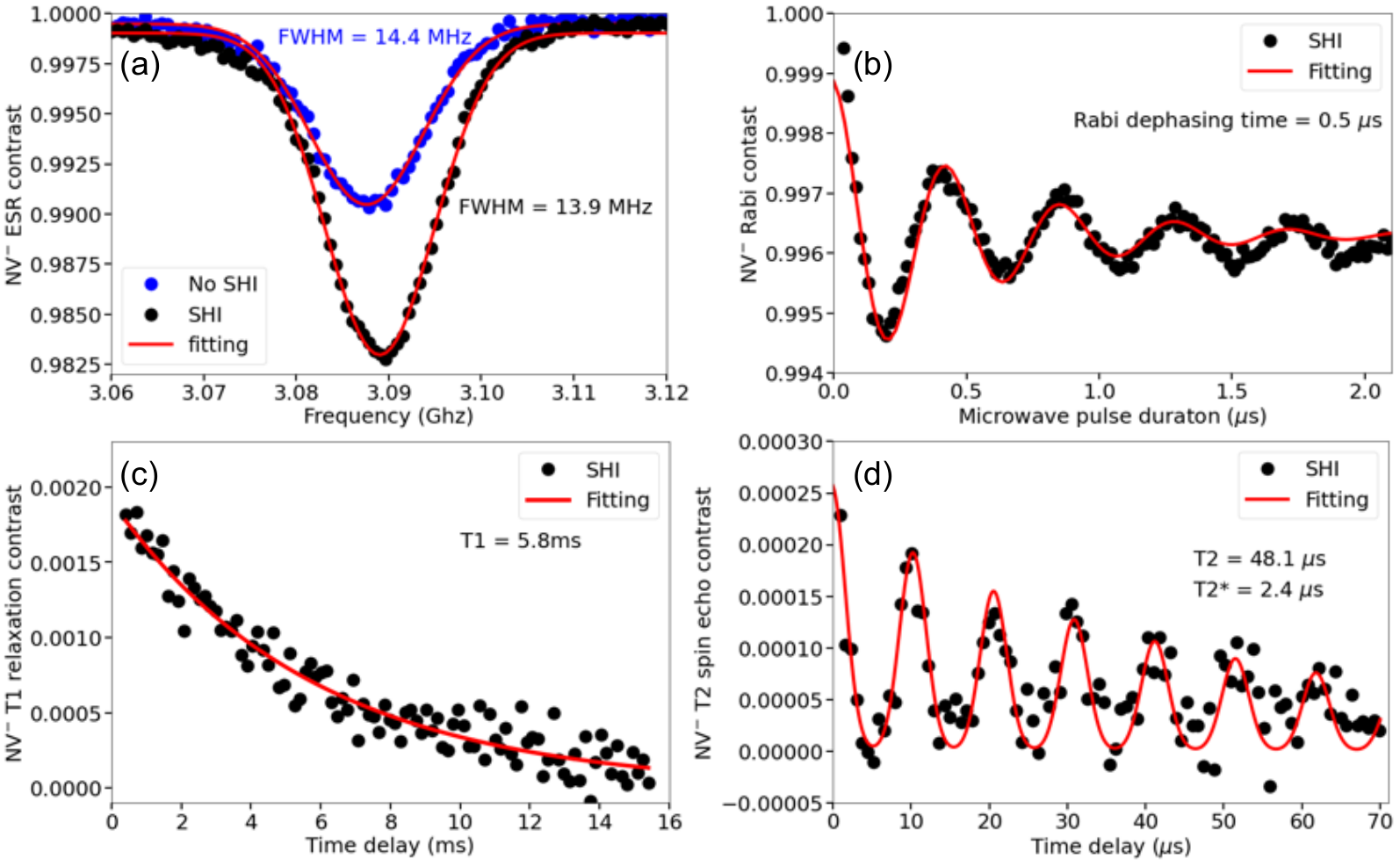}
\caption{\textbf{Electron spin properties of NV$^{-}$ centers formed by SHI characterized by ODMR.} (a) a comparison of ESR signal of the electron spin state m$_{s}$ = +1 of the NV$^{-}$ centers probed in the area with (black dots) and without (blue dots) SHI irradiation after thermal annealing, with the external magnetic field of 80 Gauss was aligned to the $<$001$>$ crystallographic orientation. Red curves correspond to the single Gaussian fitted individual resonance, which extract a FWHM of the ESR signal of 13.9 MHz and 14.4 MHz probed in the area with and supplimentary irradiation, respectively; (b) the correspondent Rabi oscillations contrast of NV$^{-}$ spins probed in the SHI irradiated area of sample B after thermal annealing. Red curve correspond to the sinusoidal function modulated by single exponential decay, which extracted Rabi dephasing time of 0.5 $\mu$s; (c) presents the correspondent T1 relaxation contrast of NV$^{-}$ spins probed in the SHI irradiated area of sample B after thermal annealing, by using a 200 ns $\pi$ pulse calibrated by the Rabi oscillation. Red curve correspond to single exponential decay, with extracted T1 coherent time of 5.8 ms; (d) presents the T2 relaxation contrast of NV$^{-}$ spins probed in the SHI irradiated area of sample B after thermal annealing, by implementing spin echo (Hahn-echo) MW pulse sequence. Red curve correspond to the fit by the model referred to Ref. \cite{shields2015efficient}, which allows to extract the T2 coherent time of 48.1 $\mu$s and T2$^{*}$ time of 2.4 $\mu$s }
\label{fig4}
\end{figure} 
Figure \ref{fig4}(a) shows a comparison of ESR signal of the electron spin state of the NV$^{-}$ centers probed in the area with (black dots) and without (blue dots) SHI irradiation. We observed that the ESR contrast of ensemble NV$^{-}$ centers formed by SHI irradiation is about twice as much of the intrinsically existed background NV$^{-}$ probed in the area of no SHI irradiation. Moreover, SHI induced ensemble of NV$^{-}$ centers presents a 13.9 MHz full width at half maximum (FWHM) of ESR signal, which is notably smaller than 14.4 MHz of background NV$^{-}$. Figure \ref{fig4} (b) shows Rabi oscillations of the SHI-induced ensemble NV$^{-} $centers \cite{bucher2019quantum}. We derived an effective dephasing time $\tau$ = 0.5 $\mu$s from the decay of the amplitude of the Rabi oscillation signal, which is comparable to dephasing times reported from NV$^{-}$ ensembles present in diamonds containing nitrogen concentration in the ppm range \cite{rubinas2018spin, mindarava2020efficient}. Since the dephasing mechanism of NV$^{-}$ is mainly governed by the spin bath \cite{mindarava2020efficient,feng2019optimizing}, the dephasing of NV$^{-}$ centers in our sample is affected by a relatively high surface nitrogen concentration and potentially by other surface noise. The depth dependent dephasing kinetics will be probed in future studies. Figure \ref{fig4} (c) presents the correspondent T1 relaxation contrast of  SHI induced ensemble of NV$^{-}$ spins, by using a 200 ns $\pi$ pulse calibrated by the prior Rabi measurement. By Fitting the experimental data with single exponential decay, we extracted a T1 coherent time of 5.8 ms. Figure \ref{fig4} (d) presents the T2 relaxation contrast of NV$^{-}$ spins probed in the SHI irradiated area of sample B after thermal annealing, by implementing spin echo (Hahn-echo) MW pulse sequence. Red curve correspond to the fit by the model referred to Ref. \cite{shields2015efficient}, which allows to extract the T2 coherent time of 48.1 $\mu$s and T2$^{*}$ time of 2.4 $\mu$s. These characterized coherent time of T1, T2 and T2$^{*}$ is even slight longer than the coherent times of background NV$^{-}$ probed in the area without SHI irradiation (as shown in supplementary Fig. S04). Overall, compared to background NV$^{-}$ center, the ensemble of NV$^{-}$ centers created by SHI presents a stronger ESR contrast, narrower linewidth, and preserves comparable T1, T2 and T2$^{*}$. It validates high fluence SHI irradiation followed with post annealing enables creating dense ensemble of NV$^{-}$ centers with superior spin properties, compared to background NV$^{-}$ centers.   

\subsection*{Single individual 1D chains of NV$^{-}$ chains formed by dilute SHI irradiation}
 We expect that our approach should be able to create single individual self-aligned quasi-1D chains of coupled NV-centers with lengths in the tens of micron range, which can be a building block for quantum information processing and they provide insights into harsh radiation-matter interactions.
 \begin{figure}[h]
\centering
\includegraphics[width=\columnwidth]{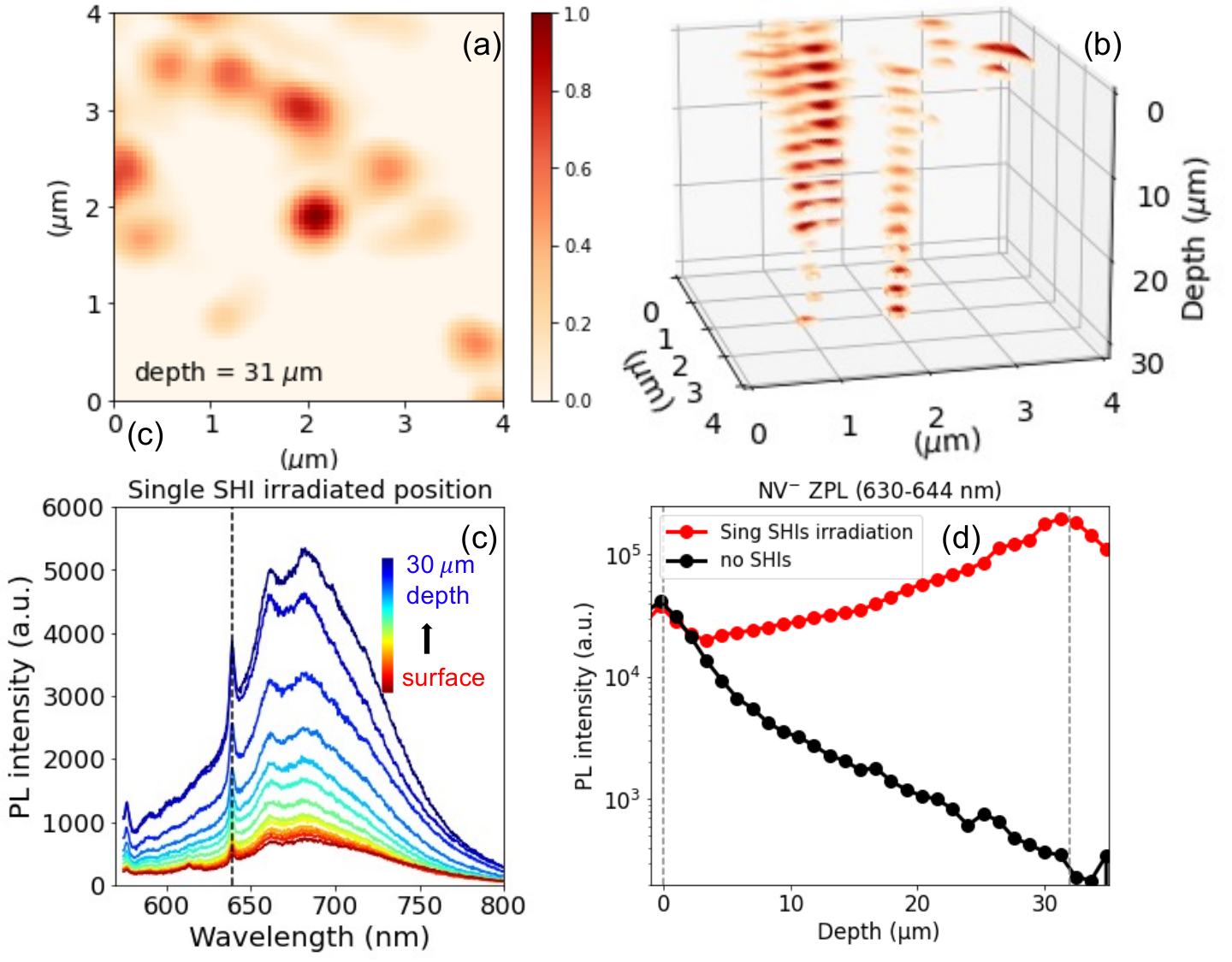}
\caption{\textbf{Optical pattern and properties of individual singe 1D NV$^{-}$ chains probed by 3D PL scan} (a) Planar view of 2D PL mapping (4$\times$4 $\mu$m$^{2}$) probed at the depth of 31 $\mu$m of the SHI irradiated area showing isolated individual 1D NV$^{-}$ chains as isolated individual bright luminescence spots; (b) Visualization of 3D luminescence patterns of individual 1D NV$^{-}$ chains reconstructed by discrete 2D PL scan probed at different depth from 0 to 30 $\mu$m; (c) color PL spectra evolution along the SHI trajectory from the sample surface to a depth of 30 $\mu$m; (d) a comparison of integrated PL intensity as a function of depth probed on (red) and off (black) the single SHI ion track.
}
\label{fig5}
\end{figure} 
 We performed 1 GeV Au ions radiation with a dilute fluence of 1$\times$10$^{8}$ ions/cm$^{2}$ on Type Ib HPHT diamond wafer (supplier: Element 6) with $<$200 ppm nitrogen impurities) at GSI UNILAC, followed by one hour post annealing under the conduction of 1000 $^{o}$C, vacuum ($<$10$^{-4}$) mbr. The yellow diamond with relative higher native nitrogen concentration is intentionally chosen for creating dense and coupled NV$^{-}$ along the SHI trajectory \cite{lake2021direct}. Subsequently, by performing 2D planar PL scan with a focused depth of 31 $\mu$m in fabricated sample, we observed isolated individual bright luminescence spots in the 2D PL mapping (4$\times$4 $\mu$m$^{2}$) probed at the SHI irradiated area, as shown in figure \ref{fig5} (a). The density of bright luminescence spot in the probe area is about 0.75$\times$10$^{8}$ cm$^{-2}$. We did not observe any of these bright luminescence spots in the area of sample where SHIs were blocked by the mask. To further verify our hypothesis, we selectively visualized in figure \ref{fig5} (b) several bright 3D luminescence pattern of individual 1D NV$^{-}$ chains, probed at the same area of figure \ref{fig5} (a). It was reconstructed by multiple discretely 2D PL scans probed at different depth from 0 to 30 $\mu$m. The isolated bright luminescence strings evidence the presence of densely coupled NV-centers created along a single ion trajectories. figure \ref{fig5} (c) reveals color center PL spectra evolution along the SHI trajectory from the sample surface to a depth of 30 $\mu$m, which comprise dominated NV$^{-}$ ZPL and phonon sideband emission and weak NV$^{0}$ component emission. Figure \ref{fig5} (d) show a comparison of integrated PL intensity as a function of depth probed on (red) and off (black) the single SHI ion track. The up to 30 µm length luminescence pattern of 1D chains of NV-centers is close to the MD simulated estimated range of isolated vacancies and defect clusters formed along ion trajectory through electronic stopping processes. Moreover, the PL intensity along the 1D NV$^{-}$ chain increase monotonically, which is line with the gradually increase of the concentration of vacancies in the form of clusters within the electron stopping range. It indicates the formation of NV-centers along SHI trajectory is governed by the interaction of native nitrogen atoms with vacancies in the form of cluster. Our optical characterization inevitably confirmed that isolated individual quasi-1D chains of NV-centers were formed by SHI irradiation and posting annealing, which leads to the conversion of native nitrogen atoms to nitrogen-vacancy centers (NV) along the ion trajectories. To realize a 1D NV spin chain for applications such as quantum registers or qubit transport channels, it is essential to precisely tailor dipole-dipole coupling between NV-centers with nanometer-scale spacing along the chain. Our prior experiment results have demonstrated\cite{lake2021direct} that the estimated conversion efficiency from nitrogen atom to NV along SHIs trajectory is approximately 15$\%$–20$\%$, assuming a nitrogen density of 100 ppm, with 1 GeV gold ion irradiation. Though this remains a rough approximation, it provides a qualitative prediction of average spacings of a few nanometers over a length exceeding ten micrometers, resulting in a quasi-1D register containing potentially over a thousand qubits. 
 
We further quantify the electron spin (m$_{s}$ = -1) properties of individual 1D NV$^{-}$ chains through ODMR. 
\begin{figure}[h]
\centering
\includegraphics[width=\columnwidth]{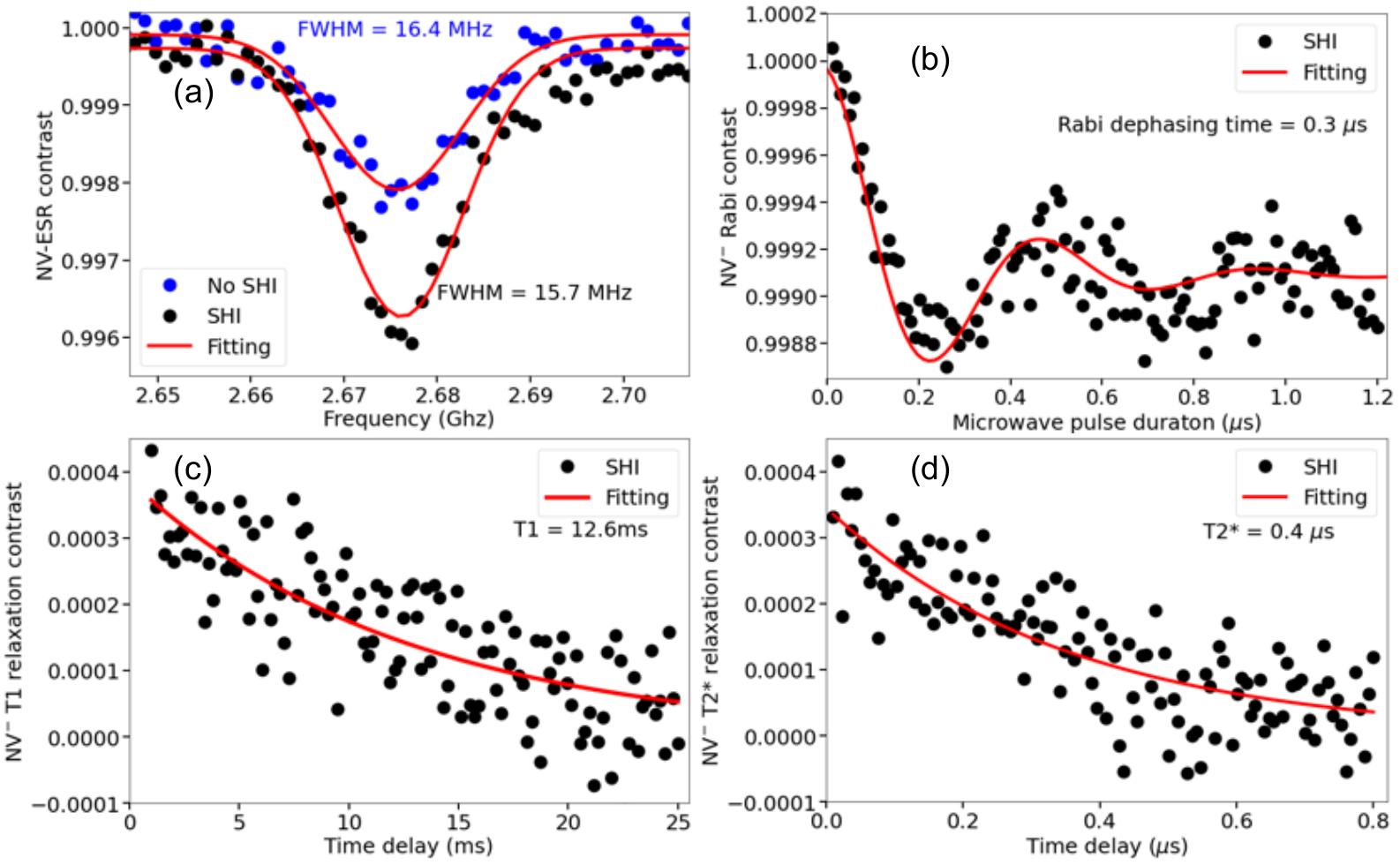}
\caption{\textbf{Electron spin properties of NV$^{-}$ centers formed by SHI characterized by ODMR.} (a) a comparison of ESR signal of the electron spin state m$_{s}$ = -1 of the NV$^{-}$ centers probed in the area with (black dots) and without (blue dots) SHI irradiation after thermal annealing, with the external magnetic field of 70 Gauss was aligned to the $<$001$>$ crystallographic orientation, and perpendicular to the 1D NV$^{-}$ chain. Red curves correspond to the single Gaussian fitted individual resonance, which extract a FWHM of the ESR signal of 15.7 MHz and 16.4 MHz  probed in the area with and without SHI irradiation, respectively; (b) the correspondent Rabi oscillations contrast of NV$^{-}$ spins probed in the SHI irradiated area of sample B after thermal annealing. Red curve correspond to the sinusoidal function modulated by single exponential decay, which extracted Rabi dephasing time of 0.3 $\mu$s; (c) presents the correspondent T1 relaxation contrast of NV$^{-}$ spins probed in the SHI irradiated area of sample B after thermal annealing, by using a 200 ns $\pi$ pulse calibrated by the Rabi oscillation. Red curve correspond to single exponential decay, with extracted T1 coherent time of 12.6 ms; (d) presents the T2 relaxation contrast of NV$^{-}$ spins probed in the SHI irradiated area of sample B after thermal annealing, by implementing spin echo (Hahn-echo) MW pulse sequence. Red curve correspond to the fit by single exponential decay, with extracted T2$^{*}$ coherent time of 0.4 $\mu$s}
\label{fig6}
\end{figure} 
An external magnetic field of 70 Gauss was aligned to the $<$001$>$ crystallographic orientation, and perpendicular to the 1D NV$^{-}$ chain. A better than 0.5 $\mu$m optical spatial resolution allows optically dressing individual 1D NV$^{-}$ chains. As shown Figure \ref{fig6} (a), individual 1D NV$^{-}$ chains show enhanced electron spin resonance contrast and narrower linewidth compared to a background of NV$^{-}$ centers that present in the nitrogen doped diamonds. This trend is in common with the comparison between the prior mentioned ensemble NV$^{-}$ centers formed by the high fluence SHI irradiation and background NV-centers. Figure \ref{fig6} (b) presents the correspondent Rabi oscillation of an individual 1D NV$^{-}$ chain, with Rabi dephasing time of 0.3 $\mu$s. Figure \ref{fig6} (c) and (d) presents the correspondent T1 (12.6 ms) and T2$^{*}$ (0.4 $\mu$s) relaxation measurement of 1D NV$^{-}$ chains. Compared to the prior mentioned ensemble NV$^{-}$ centers formed by the high fluence SHI irradiation, the 1D NV$^{-}$ chains show less ESR contrast, and shorter T1 and T2$^{*}$ coherent time. It can be due to the fact that the 1D NV$^{-}$ chains experience stronger dephasing mechanization and spin noise in the yellow diamond, which has larger nitrogen impurities ($<$200 ppm), compared the host diamond (1 ppm nitrogen impurities) of ensemble NV$^{-}$ centers. Though is beyond the scope of the study presented here, it is worthwhile to explore the approach of suppressing nitrogen impurity induced dephasing while maintain dense and coupled NV$^{-}$ centers along the 1D chain. In addition, as to the coherent properties of background NV$^{-}$ center spins probed in the area without SHI irradiation of sample D, it reveals the similar  0.3 $\mu$s Rabi dephasing time, as shown in supplementary Fig. s05. However, we did not observation of T1 and T2 relaxation contrast of by using the same measurement condition and spin readout protocol as used for 1D chain of NV$^{-}$ center. It can be due to the insufficiency signal to noise ratio of background NV$^{-}$ center to probe their spin properties. Our findings suggest the possibility that NV-centers in 1D chains with dipolar interaction between NV-centers can improve sensitivity in NV-based magnetometry applications and provide guidance on the engineering of 1D chains of NV-centers with minimal disorder for applications in quantum information processing.

\section*{Conclusions}
In conclusion, we investigated NV$^{-}$ center formation along the trajectories of swift gold and U ions in the kinetic energy range of $\sim$1 GeV in nitrogen doped diamond. By using confocal laser scanning fluorescence microscopy, we studied the interaction between native nitrogen atoms and SHI-induced vacancies during the non-equilibrium process of track formation. We also investigated thermal annealing effects on the structural and charge state conversion dynamics between GR1, NV$^{0}$ and NV$^{-}$ centers. We report strong optical emission signals from GR1-centers that are formed along the path of SHI. This direct GR1 formation by SHI through electronic stopping processes is relevant for the development of diamond-based detectors and methods for dark matter searches with candidates in the extremely high mass range. The NV-center formation dynamics by SHI can be a two-step process that is sensitive to the concentration of available nitrogen atoms around the ion trajectory. Molecular dynamics simulations of the diamond lattice subjected to energy deposition through electronic stopping processes from SHI show the formation of vacancies along SHI trajectories from the surface to the end of the ion range. Mobile vacancies can combine with nitrogen atoms in their vicinity forming NV-centers during thermal annealing in a well established NV-formation process. MD simulations indicate that both isolated vacancies and defect clusters form along the ion trajectories. Our simulations further reveal a velocity effect on the rate of vacancy formation that leads to significantly enhanced defect clustering in a depth range where the electronic stopping power is decreasing. We observed that NV$^{0}$ centers are more preferentially formed in diamond with about 1 ppm nitrogen density by SHI excitation, converting them to NV$^{-}$ centers during consecutive thermal annealing steps. This is in contrast to SHI irradiation of diamonds containing higher concentrations of nitrogen, e.g. 100 ppm, where NV-centers form during the cool down of the ion tracks \cite{lake2021direct}. The understanding of the formation mechanism of NV-centers by SHI allow us creating individual isolated quasi-1D chains of NV-centers formed by 1 GeV Au ions with a low fluence of 1$\times$10$^{8}$ ions/cm$^{2}$ radiation on the diamond with $<$200 ppm native nitrogen. The presence of densely coupled NV-centers created along a single ion trajectories average spacings of a few nanometers over a length exceeding ten micrometers, resulting in a quasi-1D register containing potentially over a thousand qubits. We probe the spin properties of the SHI-induced NV$^{-}$ centers via ODMR demonstrating that NV$^{-}$ centers formed by irradiation with SHI show competitive dephasing times,  enhanced electron spin resonance contrast, and narrower resonance lines compared to a background of NV-centers that are present in the nitrogen doped diamonds. Our findings suggest the possibility that NV-centers in 1D chains with dipolar interaction between NV-centers can improve sensitivity in NV-based magnetometry applications and provide guidance on the engineering of 1D chains of NV-centers with minimal disorder for applications in quantum information processing. The presence of NV$^{-}$ centers in quasi-1D strings along the trajectories of SHI and the formation of 2D sheets of NV$^{-}$ centers enable studies of spin textures that are complementary to those of NV$^{-}$ centers formed by more conventional methods, such as irradiation with MeV electron and protons or low energy ions. One intriguing next step is to quantify the coupling of NV-centers that are aligned along quasi-1D chains and in 2D sheets with lengths in the 20 micron range for applications as spin registers with thousands of aligned qubits.

\section*{Methods}
\subsection*{Sample preparation: SHI irradiation and thermal annealing}
Type IIa diamonds (supplier: Element 6) were used in this study with about 1 ppm nitrogen introduced during the chemical vapor deposition growth. Erbium ions were firstly implanted into diamond samples (1$\times$10$^{13}$ cm$^{-2}$ 180 keV). The pre-implanted samples were irradiated at the linear accelerator UNILAC at GSI Helmholtzzentrum (Darmstadt, Germany) using 1.1 GeV U ions (fluence of 1$\times$ 10$^{12}$ U ions/cm$^{2}$) and 0.95 GeV Au ions (2.2$\times$10$^{12}$ Au ions/cm$^{2}$). The samples were covered by a thick honey-comb shaped mask with millimeter sized openings (90\% transparency), which allows direct comparison between irradiated and nonirradiated regions during the PL measurements. To enhance the formation of NV$^{-}$ centers, the Au-ion irradiated sample B was thermally annealed in a two-step process and the optical properties were compared after the 1st and the 2nd annealing step. The 1st annealing step was for 1 hour at 800 \textdegree C in vacuum (10$^{-6}$ mbar), the 2nd annealing step was for 1 hour at 1000 \textdegree C in argon atmosphere. To create single individual quasi-1D chain of coupled NV-centers with average spacings of a few nanometers, Type Ib HPHT diamond wafer (supplier: Element 6) with $<$200 ppm nitrogen impurities) were radiated by 1 GeV Au ions radiation with a dilute fluence of 1$\times$10$^{8}$ ions/cm$^{2}$ at GSI UNILAC, followed by one hour post annealing under the conduction of 1000 $^{o}$C, vacuum ($<$10$^{-4}$) mbr.

\subsection*{Confocal photoluminescence microscopy}
Depth-resolved PL measurements were performed with a custom built confocal PL microscopy, which enables spatially resolved 3D maps of optical-active defects in semiconductor. A 100$\times$ lens with a numerical aperture (NA) of 0.95 were used to focus laser and collect PL signal. To enhanced the spatial resolution, the PL signal were refocused at the conjugate focal plane, where a pinhole with 100 $\mu$m aperture diameters is implanted to selective the centre area of excitation spot. The sample was mounted on a 3D piezo-nanopositioning sample stage. A spectrometer with 150 groves/mm grating and an electron-multiplied CCD were used to acquire the PL spectrum of color centers. The excitation wavelength of laser was 532 nm. To convert from the stage position coordinate to the actual depth where the diamond is optically excited, probe depth $z$ were refractive index corrected by multiplying the stage position $z_{stage}$ by the index of refraction in diamond, $n$=2.4, as $z = nz_{stage}$. 

\subsection*{Optical detected magnetic resonance measurement}
NV$^{-}$ center electron spin resonance and Rabi measurements were performed with a custom built platform of optical detected magnetic resonance. A 532 nm continuous wave laser with several tens of mW output power is used to excited NV$^{-}$ centers. Meanwhile, microwave (MW) pulses are generated by a digital delay/pulse generator and amplified by a high-power amplifier, and subsequently delivered to the NV-centers in diamond via a loop antenna. To implement the ESR and Rabi laser-MW pulsed sequence protocol. A PulseBlaster card synchronizes the laser and MW pulses and controls the data acquisition unit, which captures and processes the fluorescence signals to determine the NV$^{-}$ spin state. The laser is modulated by an acousto-optic modulator for timing control. NVfluorescence is collected through microscope lenses and directed to an avalanche photo diode for detection, with a band-pass filter separating the NV$^{-}$ fluorescence from the laser light. 

\section*{Data Availability Statement}
The data that support the findings of this study are available from the corresponding author upon reasonable request.

\section*{Acknowledgements}
The work is supported by the U.S. Department of Energy Office of Science, Office of Fusion Energy Sciences, under Contract No. DEAC02-05CH11231. Work at the Molecular Foundry was supported by the Office of Science, Office of Basic Energy Sciences, of the U.S. Department of Energy under Contract No. DE-AC02-05CH11231. The ODMR experiments were performed at the geoscience quantum sensing laboratory at Berkeley Lab, supported by the U.S. Department of Energy, Office of Science, Office of Basic Energy Sciences, Chemical Sciences, Geosciences, and Biosciences Division under U.S. Department of Energy Contract No. DE-AC02-05CH11231. The results presented here are based on a UMAT experiment, which was performed at the M-branch of the UNILAC at the GSI Helmholtzzentrum für Schwerionenforschung, Darmstadt (Germany) in the frame of FAIR Phase-0.

\section*{Author contributions}
T.S. and W.L., A.P and Q.J. initialized, conceived and designed the overall research. W.L. performed the depth-resolved PL characterization, ODMR measurements, and diamond thermal annealing. A.L. and C.N. performed the MD simulations and summarized the theoretical results in the manuscript with guidance from F.D.. E.B. assisted the depth-resolved PL characterization with inputs from S.A. and K.J.. M.T. and C.T. implemented the SHI irradiation. H.O.,N.A. and Z.H. supported the implementation of the ODMR measurement with inputs from R.W.. W.L. and T.S. analyzed the results with inputs from all authors. W.L. and T.S. wrote the manuscript with contribution from all authors.

\section*{Conflict of interests}
The authors have no conflicts to disclose.



\bibliography{reference} 

\end{document}